\documentclass[conference]{IEEEtran}
\IEEEoverridecommandlockouts

\usepackage{cite}
\usepackage{amsmath,amssymb}
\usepackage{graphicx}
\usepackage{booktabs}
\usepackage{array}
\usepackage{multirow}
\usepackage{xcolor}
\usepackage{tikz}
\usepackage{balance}

\usetikzlibrary{arrows.meta,positioning,fit,calc,backgrounds,shapes.misc}
\usepackage{siunitx}
\usepackage[hidelinks]{hyperref}
\usepackage{xurl}

\newcommand{\tok}[1]{\textsf{#1}}

\begin{document}

\title{Guarded Swarms in the Loop: A ROS\,2/PX4 Digital Twin for Hybrid Digital--Analog Coordination of UAV/UGV Swarms}

\author{
\IEEEauthorblockN{Uwe M. Borghoff}
\IEEEauthorblockA{\textit{Institute for Software Technology}\\
\textit{University of the Bundeswehr Munich}\\
Neubiberg, Germany\\
uwe.borghoff@unibw.de}
\and
\IEEEauthorblockN{Paolo Bottoni}
\IEEEauthorblockA{\textit{Department of Computer Science}\\
\textit{Sapienza University of Rome}\\
Rome, Italy\\
bottoni@di.uniroma1.it}
\and
\IEEEauthorblockN{Luca D'Aurizio and Remo Pareschi}
\IEEEauthorblockA{\textit{Stake Lab}\\
\textit{University of Molise}\\
Campobasso, Italy\\
l.daurizio@studenti.unimol.it\\
remo.pareschi@unimol.it}
}

\maketitle

\begin{abstract}
Swarms of unmanned aerial and ground vehicles (UAV/UGV) operating in contested environments cannot rely on software alone for physical safety, since the software stack and its networks are part of the attack surface. The \emph{Guarded Swarms} architecture addresses this by pairing a digital coordination layer, built on Topic-Based Communication Space Petri Nets (TB-CSPN), with an independent analog safety layer that can veto unsafe actuation commands. So far, the architecture has been specified and argued for, but not executed. We present the first executable instantiation of Guarded Swarms as a digital twin built on ROS\,2, the PX4 autopilot in software-in-the-loop mode, and Gazebo, with two UAVs and one UGV in a coastal counter-UAS scenario. The TB-CSPN transition chain, from multi-modal sensing through human authorization to vehicle tasking, is mapped one-to-one onto ROS\,2 nodes, and the analog layer is modeled as a per-vehicle Virtual Safety Board that gates every command. Eight groups of experiments show that the three negative guarantees of the architecture (no mission from conflicting, temporally incoherent, or unauthorized evidence) hold at three distinct stages of the chain in all 60 counterfactual runs; that threat confirmation is bounded by the narrowest sensing modality; that the safety layer exercises its full modeled envelope independently of the coordination layer; and, under fault injection, that it prevents every collision caused by a forged mission or a compromised worker once its veto is enforced by an independent gate, whereas a veto left to the compromised worker prevents none. The twin also exposes implementation-level hazards that the architectural specification does not address, and it draws an explicit line between the properties that simulation can validate and those that only hardware can.
\end{abstract}


\section{Introduction}

Swarms of unmanned vehicles have moved from single teleoperated platforms to coordinated groups that share a mission, with clear benefits in coverage, redundancy and resilience~\cite{brambilla2013swarm,chung2018aerial}. Coordination in such swarms is almost entirely a software matter: planners, middleware and distributed controllers decide who does what and when. In contested environments this is also a weakness. An adversary can spoof navigation or sensing signals~\cite{kerns2014spoofing}, inject commands into a compromised stack, or deny communications; and even without an adversary, a software defect or a perception error can drive a vehicle into a physically dangerous state. If the software layer is the only safeguard, nothing downstream prevents an unsafe command from reaching the motors.

The \emph{Guarded Swarms} architecture~\cite{borghoff2025guardedTD,borghoff2026guardedFI,borghoff2026sensor} answers this problem by splitting responsibility between two layers that do not trust each other. A \emph{digital coordination layer}, formalized in the TB-CSPN framework~\cite{borghoff2025tbcspn}, turns heterogeneous sensor evidence into authorized missions through explicit, auditable Petri-net transitions. An \emph{analog safety layer}, present on each vehicle and independent of software, validates or vetoes actuation commands at the last moment before they reach the actuators. The two layers share no state and meet only at the actuation command, where the safety layer has the last word.

Both the sensor-to-mission side~\cite{borghoff2026sensor} and the digital--analog interface~\cite{borghoff2025guardedTD,borghoff2026guardedFI} of the architecture have been published as architectural contributions, with quantitative validation explicitly deferred to future work. This paper takes that step. Its contributions are:
\begin{enumerate}
  \item \emph{An executable instantiation} of Guarded Swarms as a digital twin~\cite{grieves2017dt} on ROS\,2~\cite{macenski2022ros2}, PX4 in software-in-the-loop (SITL) mode~\cite{meier2015px4} and Gazebo~\cite{koenig2004gazebo}, in which each TB-CSPN transition is a ROS\,2 node, each token type is a typed message, and the analog layer is a per-vehicle \emph{Virtual Safety Board} (VSB) that gates every command.
  \item \emph{An evaluation protocol} in eight groups that measures latency, negative guarantees, modality dependence, dispatch fan-out, safety-layer behavior, resilience to fault injection and sensitivity to timing parameters on the coastal counter-UAS case study of~\cite{borghoff2026sensor}.
  \item \emph{Lessons from execution}: implementation hazards that silently defeat or degrade architectural guarantees (clock-domain mismatches and manipulable time bases, non-idempotent re-firing of governed release, a veto enforced inside the component it is meant to restrain), an emergent distinction between mission validity and track persistence, and an explicit account of which guarantees the twin validates and which remain for hardware.
\end{enumerate}

The work originates in the bachelor's thesis of the third author~\cite{daurizio2026thesis}. The architecture is not a contribution of this paper; its executable realization and the evidence it yields are.

\section{Military Rationale}
\label{sec:military}

\emph{Operational problem.} Small uncrewed aircraft are a persistent threat to coastal and critical installations such as ports, naval bases and offshore energy infrastructure. They fly low and slow, have small signatures, and approach from directions that no single sensor covers well, so detection depends on fusing radar, passive RF, acoustic and EO sensing under time pressure~\cite{salatino2026interplay,samaras2019cuas,taha2019drone}. Uncrewed interceptors and ground vehicles offer coverage and endurance that crewed assets cannot sustain, but they operate exactly where adversaries use electronic warfare: GNSS spoofing~\cite{kerns2014spoofing}, jamming, and intrusion into the software and links that coordinate them. The coastal counter-UAS case study used in this paper is a compact instance of this problem.

\emph{Requirements.} Three requirements follow, and each maps onto a specific mechanism of Guarded Swarms. (i)~\emph{Human authority under rules of engagement.} No uncrewed response may be launched without an accountable human decision, and the decision must be bound to the specific threat and to a limited time. In TB-CSPN, authorization is a coordination transition (T6) that combines a confirmed threat with a time-bounded supervisor token; the ``authorization denied'' counterfactual is precisely the case of rules of engagement that do not permit a response. (ii)~\emph{Resilience under compromise and degradation.} When software or links are compromised, each vehicle must still refuse physically unsafe commands. This is the role of the analog safety layer, which guarantees local safety deterministically while global coordination degrades gracefully. (iii)~\emph{Accountability and assurance.} Commanders and accreditation authorities must be able to reconstruct why an action was taken, and autonomous functions must be tested, evaluated, verified and validated before fielding~\cite{nato2021ai,nato2024ai,dod3000}. 
Provenance chains from command to observation and to the authorizing supervisor address the former; the digital twin addresses the latter.

\emph{Why a digital twin.} Live trials of autonomous swarm behavior are expensive, constrained by range safety, and unable to stage many of the cases that matter most, such as a confirmed threat with authorization withheld, evidence that arrives just too late, or a sensor lost in mid-engagement. A twin that runs the real autopilot code allows these cases to be rehearsed systematically and repeated, and it turns architectural claims into measurable properties. It also produces operational findings: in our scenario, it showed that threat confirmation is bounded by the coverage of the EO sensor, which is a sensor-placement result as much as a software one. We see the twin as a test, evaluation, verification and validation (TEV\&V) instrument that precedes, but does not replace, hardware-in-the-loop and field trials.

\section{Background and Related Work}

\subsection{TB-CSPN and Guarded Swarms}

TB-CSPN~\cite{borghoff2025tbcspn} separates a \emph{semantic} layer, which interprets context and may be realized by language models, planners or specialized AI modules, from a \emph{coordination} layer realized as colored Petri nets~\cite{jensen2009cpn} structured as workflow nets. Participants take one of three roles: \emph{supervisors} (human) apply policy and enable authorization transitions; \emph{consultants} interpret observations into semantic tokens; \emph{workers} translate mission tokens into platform commands without mission-level reasoning. Tokens are typed, time-stamped and carry a time-to-live (TTL); transitions that fuse tokens from different sources fire only if their validity windows overlap within a bound, so that decisions rest on temporally coherent evidence. The resulting \emph{bounded non-determinism} lets several transitions be enabled at once while confining the admissible evolutions by token availability, guards and timing.

Guarded Swarms~\cite{borghoff2025guardedTD,borghoff2026guardedFI} extends TB-CSPN to the physical domain. Each vehicle carries an analog safety perimeter made of independent enforcement channels: window-comparator thrust limiters, gyroscope-based attitude and rate trips, IR/ultrasonic proximity ``bumpers'', power and thermal protection, an RF guard tone, and a physical emergency stop. The channels act in parallel under OR-inhibition logic: any violated guard can unilaterally block or clamp the command; if none fires, the command passes unchanged. The authors read this as contract-based design~\cite{benveniste2018contracts}: the digital layer may vary, the analog layer imposes a fixed safety contract. The companion paper~\cite{borghoff2026sensor} specifies the sensor-driven mission synthesis chain for mixed UAV/UGV swarms and a coastal-surveillance case study with three counterfactual trajectories, which we reproduce here. A hardware realization path for the enforcement layer is being pursued separately~\cite{borghoff2026fpga}.

\subsection{Related Work}

\emph{Runtime assurance.} The Simplex architecture~\cite{sha2001simplicity} protects a complex controller with a simple verified one, and the idea has been carried into robotics runtime assurance frameworks such as SOTER~\cite{desai2019soter}, into verified runtime monitors for cyber-physical models such as ModelPlex~\cite{mitsch2016modelplex}, and into UAS practice through ASTM F3269~\cite{astmF3269}. These approaches share with Guarded Swarms the idea of a guarantee layer separated from the main logic, but realize it in software, which leaves the guarantee within the attack surface. Guarded Swarms moves the last line of defense outside the digital domain.

\emph{Multi-sensor detection.} Counter-UAS detection fuses radar, RF, acoustic and electro-optical (EO) sensing~\cite{samaras2019cuas,taha2019drone}, and Bayesian fusion has been applied to heterogeneous maritime sensing~\cite{gaglione2020fusion}. The digital layer considered here proposes no new fusion algorithm; its contribution is a coordination contract that prevents the fusion of stale or conflicting evidence and makes every mission traceable to its observations.

\emph{Human oversight and trust.} NATO's principles of responsible use of AI call for responsibility and accountability, explainability and traceability, reliability, and governability~\cite{nato2021ai}; national policy similarly requires appropriate human judgment over autonomous functions~\cite{dod3000}, and the philosophical literature frames this as meaningful human control~\cite{santoni2018mhc}. 
Recent studies illustrate both the benefits and the limitations of
AI-supported analysis. In an experiment with military intelligence analysts,
AI assistance improved assessments under time pressure and produced probability
estimates closer to those of experts, but did not increase confidence in the
underlying sources and remained limited when the available information was
ambiguous or contradictory~\cite{NitzlCKB25}. Complementing these findings,
Hermanns and Teubner found that time pressure shifts the evaluation of AI
outputs towards heuristic processing and reduces users' ability to distinguish
correct from faulty responses, particularly when tasks are complex or the AI
provides unreliable advice~\cite{hermanns2026under}. Moreover, Papaioannou
showed that LLM-generated accounts of contested historical events are not
neutral: the language of the prompt can alter terminology, causal
interpretation, the attribution of responsibility, and the inclusion or
omission of competing perspectives~\cite{Papaioannou26}. Taken together, these
results indicate that AI can accelerate and improve analytical work, but that
effective human--AI collaboration still requires meaningful human judgment,
critical scrutiny, and control, especially when evidence is uncertain,
contested, or open to multiple interpretations.
In TB-CSPN, this requirement is realized by representing human authorization
as an integral coordination transition rather than as an external approval
step: authorization is explicitly bound to the relevant threat, propagated
through the provenance chain, and required before any mission can be released.
Earlier work by some of the authors addressed accountability in
semi-centralized swarms through centralized monitoring and
blockchain~\cite{carovilla2023}; the resulting dependence on a single monitoring
point motivated the transition to distributed coordination spaces and,
ultimately, to TB-CSPN.

\emph{Trust and assurance in emerging swarm architectures.}
Recent work addresses complementary forms of trust in autonomous swarms.
Lassfolk and Kari~\cite{Lassfolk2026} propose a certificate-based, two-layer
trust model and onboarding procedure for heterogeneous coalition drone swarms.
Their conceptual framework governs whether a platform may enter the swarm and
exchange information, whereas Guarded Swarms governs how admitted observations,
authorizations, and tasks may produce physical action. SHIELD
\cite{WasifMYLKNC26} combines evidential uncertainty, dynamic inter-agent
trust, intrusion response, and event-triggered human escalation in simulated
swarms exposed to several cyber-physical attacks. These approaches strengthen
identity, behavioral trust, and compromise detection within the digital
domain. They are therefore complementary to, rather than substitutes for, an
actuation-level guard that remains able to reject an unsafe command when the
coordination software itself cannot be trusted.

\emph{Operational swarm platforms and programs.}
Recent defense developments show that software-defined coordination of
heterogeneous uncrewed systems is moving from experimental prototypes toward
field-oriented capabilities. DARPA's completed \textsc{OFFSET} program
established an influential precursor by combining an open swarm architecture,
human--swarm interaction and repeated live experiments with large mixed
UAV/UGV formations\footnote{\url{https://www.darpa.mil/research/programs/offensive-swarm-enabled-tactics}}.
More recent systems place a common mission-autonomy layer above heterogeneous
platforms. Auterion's \textsc{Nemyx}, for example, has been demonstrated with
fixed-wing and FPV aircraft from different manufacturers operating under shared
mission logic\footnote{\url{https://auterion.com/auterion-performs-worlds-first-multi-manufacturer-swarm-strike-demonstration/}},
while Anduril's \textsc{Lattice for Mission Autonomy} is intended to coordinate
multi-domain teams of government-owned, third-party and proprietary uncrewed
systems\footnote{\url{https://www.anduril.com/article/special-operations-command-selects-anduril-as-mission-autonomy-systems-integration-partner/}}.
Helsing similarly couples its software-defined \textsc{HX-2} aircraft to the
\textsc{Altra} platform for AI-supported mission planning, sensor fusion and
coordinated employment under human oversight\footnote{\url{https://helsing.ai/hx-2}}.
At program level, the UK Software Defined Swarms Concept Demonstrator has
tested competing industry implementations in realistic environments and
delivered an eight-UAV capability test bed for continued Army
experimentation\footnote{\url{https://www.gov.uk/government/news/dstl-drone-swarm-accelerates-army-autonomy-ambition}}.
These developments confirm the operational importance of scalable mission
orchestration, platform interoperability and human supervision. Publicly
available descriptions, however, provide little evidence of formally specified
coordination semantics or of a physically independent enforcement boundary
that can veto unsafe commands after compromise of the mission software. The
present work addresses this complementary assurance problem rather than
proposing another swarm-control or mission-planning algorithm.

\emph{Simulation-based validation.} ROS\,2/Gazebo pipelines are standard for validating multi-robot coordination before deployment, and PX4 SITL runs the same low-level control code used on real autopilots. 
Recent work has gone further by coupling formal coordination models to
executable twins: a bigraph-based framework for multi-UAV landing combines
model checking, ROS~2 execution and cyber--physical synchronization to enforce
conflict-free pad allocation~\cite{zhang2025bigraph}. Its focus is verified
landing management rather than mission synthesis under contested sensing or
independent actuation-level enforcement.
We use this stack precisely because it lets the vehicle dynamics be credible while the architecture-specific layers remain under full experimental control.

\section{The Digital Twin}

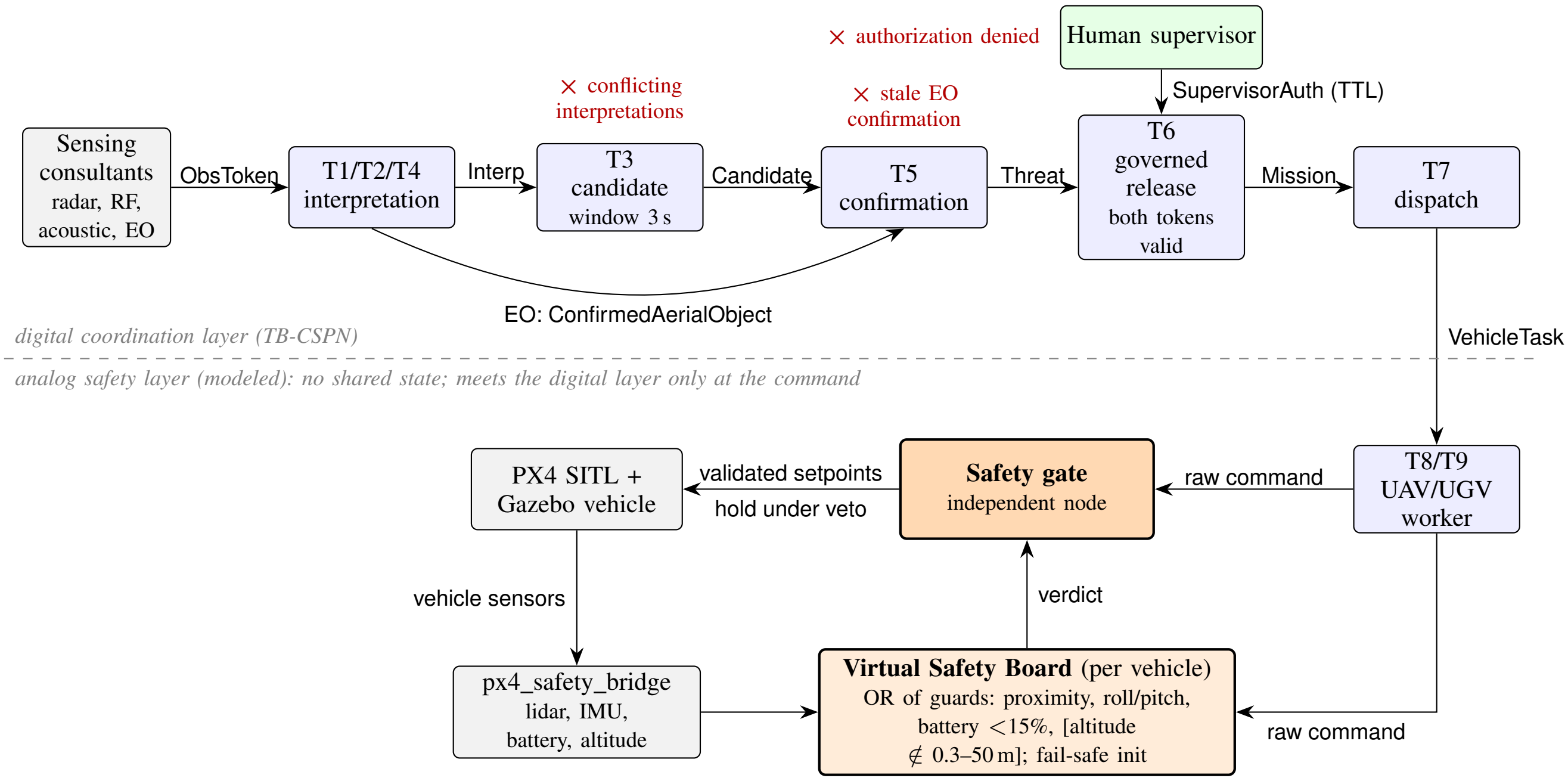
\begin{figure*}[t]
\centering
\begin{tikzpicture}[
  font=\footnotesize,
  >={Stealth[length=2mm]},
  tr/.style={draw, rounded corners=2pt, fill=blue!7, minimum height=9mm, align=center, inner sep=2pt, text width=17mm},
  ag/.style={draw, rounded corners=2pt, fill=gray!10, minimum height=9mm, align=center, inner sep=2pt},
  sb/.style={draw, thick, rounded corners=2pt, fill=orange!15, minimum height=11mm, align=center, inner sep=3pt},
  hu/.style={draw, rounded corners=2pt, fill=green!10, minimum height=7mm, align=center, inner sep=2pt},
  lbl/.style={font=\scriptsize\sffamily, midway, above, inner sep=1pt},
  blk/.style={font=\scriptsize, text=red!70!black, align=center},
  node distance=6mm and 6mm
]
\node[ag, text width=15mm] (sens) {Sensing\\consultants\\{\scriptsize radar, RF,}\\{\scriptsize acoustic, EO}};
\node[tr, right=13mm of sens] (interp) {T1/T2/T4\\interpretation};
\node[tr, right=9mm of interp] (t3) {T3\\candidate\\{\scriptsize window 3\,s}};
\node[tr, right=13mm of t3] (t5) {T5\\confirmation};
\node[tr, right=10mm of t5] (t6) {T6\\governed release\\{\scriptsize both tokens valid}};
\node[tr, right=12mm of t6] (t7) {T7\\dispatch};
\node[hu, above=5mm of t6] (sup) {Human supervisor};

\draw[->] (sens) -- node[lbl]{ObsToken} (interp);
\draw[->] (interp) -- node[lbl]{Interp} (t3);
\draw[->] (t3) -- node[lbl]{Candidate} (t5);
\draw[->] (t5) -- node[lbl]{Threat} (t6);
\draw[->] (t6) -- node[lbl]{Mission} (t7);
\draw[->] (sup) -- node[right, font=\scriptsize\sffamily]{SupervisorAuth (TTL)} (t6);
\draw[->] (interp.south) to[bend right=25] node[below, font=\scriptsize\sffamily]{EO: ConfirmedAerialObject} (t5.south);

\node[blk, above=1mm of t3] {\(\boldsymbol\times\) conflicting\\interpretations};
\node[blk, above=1mm of t5] {\(\boldsymbol\times\) stale EO\\confirmation};
\node[blk, left=1mm of sup] {\(\boldsymbol\times\) authorization denied};

\node[tr, below=24mm of t7] (wk) {T8/T9\\UAV/UGV worker};
\node[sb, left=22mm of wk, text width=26mm, fill=orange!30] (gate) {\textbf{Safety gate}\\{\scriptsize independent node}};
\node[ag, left=24mm of gate, text width=22mm] (px4) {PX4 SITL +\\Gazebo vehicle};
\node[sb, below=12mm of gate, text width=44mm] (vsb) {\textbf{Virtual Safety Board} (per vehicle)\\{\scriptsize OR of guards: proximity, roll/pitch,}\\{\scriptsize battery $<$15\%, [altitude $\notin$ 0.3--50\,m]; fail-safe init}};
\node[ag, text width=26mm] (br) at (px4 |- vsb) {px4\_safety\_bridge\\{\scriptsize lidar, IMU, battery, altitude}};

\draw[->] (t7) -- node[right, font=\scriptsize\sffamily]{VehicleTask} (wk);
\draw[->] (wk) -- node[lbl]{raw command} (gate);
\draw[->] (gate) -- node[lbl, align=center]{validated setpoints} node[below, font=\scriptsize\sffamily, midway]{hold under veto} (px4);
\draw[->] (wk.south) |- node[pos=0.75, below, font=\scriptsize\sffamily]{raw command} (vsb.east);
\draw[->] (vsb.north) -- node[right, font=\scriptsize\sffamily]{verdict} (gate.south);
\draw[->] (px4.south) -- node[left, font=\scriptsize\sffamily]{vehicle sensors} (br.north);
\draw[->] (br) -- (vsb);

\coordinate (sepL) at ($(sens.west)+(-2mm,-19mm)$);
\coordinate (sepR) at ($(t7.east)+(2mm,-19mm)$);
\draw[dashed, gray] (sepL) -- (sepR);
\node[font=\scriptsize\itshape, gray, anchor=south west] at (sepL) {digital coordination layer (TB-CSPN)};
\node[font=\scriptsize\itshape, gray, anchor=north west] at (sepL) {analog safety layer (modeled): no shared state; meets the digital layer only at the command};
\end{tikzpicture}
\caption{Executable Guarded Swarms. Top: the TB-CSPN chain, each transition a ROS\,2 node; red marks show where the three counterfactuals are blocked. Bottom: the command-level gate on each vehicle. The worker proposes a raw command; the Virtual Safety Board, fed only by the vehicle's own sensors, validates or vetoes it; an independent safety gate between worker and autopilot forwards only validated commands.}
\label{fig:arch}
\end{figure*}

\subsection{Platform}

The twin is organized in seven ROS\,2 packages that mirror the architectural separation: message definitions (tokens and commands), sensing consultants, TB-CSPN transitions, workers and intruder control, safety (VSB and PX4 bridge), launch files, and Gazebo worlds and models. It runs on Ubuntu 22.04 with ROS\,2 Humble, Gazebo Harmonic 8.11, PX4-Autopilot SITL at a pinned commit, the Micro-XRCE-DDS agent, and the \texttt{ros\_gz} bridge. The two UAVs are PX4 \texttt{x500} quadrotors extended with a forward lidar and an EO camera; the UGV is the PX4 \texttt{r1\_rover} extended with a lidar. Sensors are added through model overlays, leaving PX4 sources untouched. All experiments ran on a single workstation (Intel Core i5-12400, 32\,GB RAM, NVIDIA RTX\,3050). The system is launched in two stages: the physical layer (Gazebo, three PX4 instances in lockstep, safety bridges and VSBs) is brought up first, then the coordination layer (sensing, TB-CSPN chain, intruder). All timing parameters, modality enable flags, the supervisor decision, and the intruder entry vector are launch arguments, so every experimental configuration is obtained without code changes.

\subsection{Coordination Layer as an Executable Net}
\label{sec:twin-net}

Fig.~\ref{fig:arch} shows the chain. Four sensing consultants (radar, RF, acoustic, EO) synthesize single-modality observations from simulation ground truth according to each modality's coverage profile, with detection probability below one and confidence decaying with distance, and publish them as \tok{ObsToken}s carrying source, confidence, timestamp and TTL. A single parametric interpretation node, instantiated per modality, realizes T1 (radar $\to$ \tok{PossibleUAV}), T2 (RF $\to$ \tok{LikelyDroneLink}) and T4 (EO $\to$ \tok{ConfirmedAerialObject}). T3 fires a \tok{CandidateTarget} only on \emph{concurring} radar and RF interpretations within a 3\,s synchronization window; T5 promotes a candidate to \tok{ConfirmedThreat} when a valid EO confirmation joins it within 3\,s. T6 releases a \tok{MissionToken} only when a valid \tok{ConfirmedThreat} and a valid \tok{SupervisorAuth} for the same target are both valid at the same instant. T7 dispatches an intercept task to the UAV nearest to the target and an observation task to the UGV; workers T8 and T9 translate tasks into trajectory or drive setpoints, and the intercepting UAV then tracks the threat estimate autonomously as new confirmations arrive. Interpretation and candidate tokens have a 4\,s TTL, threats and missions a 10\,s TTL, and authorizations a 15\,s TTL. These values are not prescribed by the architecture; they were chosen to absorb asynchronous arrivals while rejecting stale evidence.

Three implementation decisions are worth stating, because each carries an architectural property into code.

\emph{Timestamp inheritance.} An interpretation token inherits the validity instant of the observation it derives from, not the instant at which it is produced. Windowed joins therefore operate on the true age of the evidence. Without this, a late EO confirmation would look fresh, and the stale-evidence counterfactual could not be reproduced faithfully. In the current implementation this holds for interpretations only: T3 and T5 stamp the tokens they produce with the instant of emission. A confirmed threat built on EO evidence up to about 2\,s old therefore receives a full 10\,s validity, as the fusion-window sweep shows (Sec.~\ref{sec:sweeps}). Propagating the age of the oldest supporting evidence to candidates and threats is a simple correction, deferred because it would change all the measurements reported here.

\emph{Governed release.} The authorization token is not consumed by T6: while it remains valid and the threat keeps being reconfirmed, T6 may re-emit a mission for the same target. In Petri-net terms, the authorization behaves like a test (read) arc with a TTL. Release is therefore a governed state that lasts as long as the authorization does, rather than a single event.

\emph{Provenance.} Every interpretation token carries an \texttt{origin} chain that propagates to the final \tok{VehicleTask}, so that any executed command can be traced back through mission, authorization, confirmed threat and intermediate interpretations to raw observations, and to the supervisor who authorized it.

\subsection{Analog Layer as a Virtual Safety Board}

In hardware, the safety layer is a set of analog circuits. In the twin it is an independent ROS\,2 node per vehicle, fed by a bridge that extracts lidar proximity, attitude, battery state and altitude from the PX4/Gazebo stack. It shares no state with the coordination nodes and subscribes to no token topic. At every control cycle, the worker publishes its proposed command as a \emph{raw} command, and the VSB evaluates its guards under OR logic and publishes a verdict. In the first implementation the worker itself applied the verdict, holding position (UAV) or zeroing thrust (UGV) under veto. Fault injection (Sec.~\ref{sec:fault}) showed that this placement is not sufficient, because a compromised worker can simply ignore the verdict. The twin therefore adds an independent \emph{safety gate} per vehicle: a separate node, interposed between the worker and PX4, that forwards a command only if the VSB has validated it. With the gate in place, no component of the coordination layer, workers included, has a path to the actuators that bypasses the VSB.

The \emph{core} envelope covers the channels observable in simulation: proximity ($<$1.5\,m), attitude (roll or pitch $>$30\textdegree) and battery ($<$15\%, a proxy for power protection that does not model current or temperature limits). Thresholds are per-vehicle parameters, to be chosen from each vehicle's speed and stopping distance; the values above are the board's defaults, kept on the UGV. In the fault-injection runs (Sec.~\ref{sec:fault}) the UAV boards veto below 10\,m and release above 12\,m, because under veto a UAV holds position but still has to brake: approaching at 5\,m/s, the default PX4 cruise speed, it stops about 4\,m after crossing the threshold, so a 1.5\,m threshold would measure braking rather than the architecture. For the same reason their attitude limit is 60\textdegree: PX4 tilts a multirotor up to 45\textdegree\ when accelerating, and treats attitudes beyond 60\textdegree\ as a failure. An \emph{extended} envelope adds an altitude perimeter (0.3--50\,m), which is not one of the published channels but shows that new guards can be added without changing enforcement logic. Thrust clamping, thermal protection, the RF guard tone and the opto-isolated emergency stop require dedicated hardware and are not modeled. A sensor reading older than 2\,s is treated as stale, and the board starts in a fail-safe state, vetoing until every sensor has delivered a valid reading.

\subsection{Scenario}

The Gazebo world is a compressed 400$\times$400\,m stretch of coast, fronting the sea. Two UAVs hover symmetrically about the intruder's nominal entry axis at 12\,m and 20\,m (vertical separation); the UGV is placed further back, near the shoreline. A small aerial intruder enters from the sea at a constant 6\,m/s after an 8\,s start delay. Its motion is commanded open-loop as a function of time, which makes trajectories identical across runs. Radar and RF have wide coverage over the sea; acoustic coverage is limited to the shoreline; the EO camera is fixed on one UAV with a narrow forward cone (azimuth 90\textdegree, tilt $-$20\textdegree, half-angle 30\textdegree, range 120\,m). In the nominal run, radar and RF raise a candidate, the EO camera confirms it as the intruder enters its cone, the supervisor authorizes, one UAV intercepts and the UGV observes.

\section{Evaluation}

All times are in simulation time, so that measurements remain comparable when the real-time factor drops below one under the load of three lockstepped PX4 instances. Measurements are extracted offline from ROS\,2 bag recordings. Binary outcomes are reported with 95\% Wilson intervals; latencies as mean, median, standard deviation and range. The eight groups address seven questions: (Q1) how long the chain takes and where the time goes (Groups 1--2); (Q2) whether unauthorized or incoherent missions are blocked (Group 3); (Q3) how detection depends on geometry and modalities (Group 4); (Q4) how dispatch traffic grows with fleet size (Group 5); (Q5) whether the safety layer works independently of coordination (Group 6); (Q6) whether it holds when the digital layer is compromised (Group 7); (Q7) how sensitive the chain is to authorization delay and fusion window (Group 8). For Groups 1--4, each of the 11 configurations was run 20 times, with Gazebo, PX4 and all nodes restarted from scratch for every run. The runs were driven by scripts at a tagged version of the code. Run $n$ uses the same random seed in every configuration, so that configurations are compared on the same noise draws, and configurations are interleaved rather than run in blocks. Every run passes an automated cleanliness check: the simulated clock reaches all nodes, a single intruder source is present, and the intruder moves at its nominal 6.00\,m/s in simulation time. Of 224 attempts, four were repeated (two interrupted by hand, one in which the scenario did not start, one in which no simulated clock reached the nodes), leaving 220 valid runs, all clean. Authorization is issued by a dedicated supervisor node with a configurable delay, set to zero in these runs, so that the measured latency is that of the architecture alone.

\subsection{Nominal Chain and Latency (Q1)}

All 20 nominal runs formed a mission (95\% CI 83.9--100\%); with the frontal-entry configuration, which is the same scenario, 40/40 (91.2--100\%). Every \emph{authorize} decision was followed by a mission (60/60 across the three configurations that issue one, 94.0--100\%), so T6 always received the decision. Table~\ref{tab:latency} decomposes the latency.

\begin{table}[t]
\centering
\caption{Chain latency, nominal scenario (simulation seconds, 20 runs).}
\label{tab:latency}
\setlength{\tabcolsep}{4pt}
\begin{tabular}{@{}lrrrc@{}}
\toprule
Stage & Mean & Median & SD & Range \\
\midrule
Observation $\to$ Candidate        & 11.07 & 11.16 & 0.25 & 10.47--11.37 \\
Candidate $\to$ Confirmed threat   & 1.09  & 1.05  & 0.10 & 1.00--1.20 \\
Confirmed threat $\to$ Mission$^{a}$ & 0.00 & 0.00 & 0.00 & -- \\
Mission $\to$ Vehicle task         & 0.00  & 0.00  & 0.00 & -- \\
\midrule
End-to-end (obs.\ $\to$ mission)    & 12.16 & 12.23 & 0.25 & 11.50--12.43 \\
\bottomrule
\end{tabular}\\[2pt]
\footnotesize $^{a}$Authorization delay set to zero; human deliberation time is excluded by design.
\end{table}

With the authorization delay set to zero, the end-to-end latency is the time the architecture needs to assemble concordant evidence: 12.2\,s on average, with a standard deviation of 0.25\,s over 20 runs. This time is set by the scenario geometry, not by processing. The first observation comes from the acoustic sensor, which does not take part in T3; T3 requires radar and RF. RF, which has the longest range, first detects the intruder 6.77\,s after the acoustic sensor, radar 10.87\,s after it, and EO 12.01\,s after it. An audit of earlier recordings showed that these first detections occur at the configured coverage limits. The candidate follows the first radar detection by exactly 0.20\,s in every run, which equals the modeled radar latency, while the processing time of T3 stays below one millisecond. Promotion to confirmed threat takes $1.09\pm0.10$\,s, and dispatch after release is immediate within measurement resolution. Because latencies are measured from the first observation of any modality, they depend on which sensor sees the intruder first. Without acoustic sensing, the reference moves to the first RF detection and the end-to-end figure drops to 5.38\,s, but the chain itself is unchanged: from first RF detection to candidate takes 4.32\,s, against 4.30\,s in the nominal runs. The architecture guarantees that no action precedes authorization; it does not, and should not, bound how fast authorization arrives.

\subsection{Negative Guarantees (Q2)}

The three counterfactuals of~\cite{borghoff2026sensor} were run 20 times each (Table~\ref{tab:negative}). In all 60 runs no mission was formed, and in each case the block occurred at a different stage of the chain. With \emph{conflicting interpretations}, the RF consultant is forced to label the contact \tok{LikelyCivilianTraffic}; the concordance guard of T3 is not satisfied and not even a candidate forms, although the EO sensor confirms an aerial object in every run. With a \emph{stale EO confirmation}, a controlled delay pushes the EO token outside the window; the candidate and the EO confirmation form in all 20 runs, but T5 never fires. With \emph{authorization denied}, evidence accumulates normally (on average 43 \tok{ConfirmedThreat} tokens per run) and the supervisor node issues a \emph{deny} decision, but T6 never fires: the system recognizes the threat and does not act. The threat remains in the coordination space for inspection or later authorization. In no case is a central arbiter involved: each block follows from a transition guard or a timing constraint.

\begin{table}[t]
\centering
\caption{Negative guarantees (20 runs per counterfactual).}
\label{tab:negative}
\setlength{\tabcolsep}{4pt}
\begin{tabular}{@{}llcc@{}}
\toprule
Counterfactual & Blocked at & Missions & 95\% CI \\
\midrule
Conflicting interpretations & T3 (candidate) & 0/20 & 0--16.1\% \\
Stale EO confirmation       & T5 (confirmation) & 0/20 & 0--16.1\% \\
Authorization denied        & T6 (release) & 0/20 & 0--16.1\% \\
\bottomrule
\end{tabular}
\end{table}

\subsection{Geometry and Modality Dependence (Q3)}

Table~\ref{tab:modalities} reports two experiments, with 20 runs per configuration. With a frontal entry, all four modalities detect the intruder and the threat is confirmed in every run. From a deep-east or a rear entry, radar and RF still detect the intruder and a candidate forms in every run, but the fixed EO cone never frames it, so the threat is never confirmed; from the rear, acoustic coverage is lost as well. The candidate forms earlier from the deep east, where radar sees the intruder first ($6.72\pm0.14$\,s after first observation), and later and with more dispersion from the rear ($14.66\pm1.00$\,s). The ablation shows a clear hierarchy: radar and RF are each necessary for the candidate, EO is necessary for confirmation, and acoustic sensing is redundant for the chain as configured. Detection is therefore \emph{directional}: wide-coverage modalities establish presence, while confirmation is bounded by the narrowest modality. The coordination logic behaves correctly here, refusing to confirm without visual evidence; the limitation lies in sensor placement, which the twin makes measurable.

\begin{table}[t]
\centering
\caption{Candidate formation and mission by entry vector and by disabled modality (20 runs each; 20/20 has 95\% CI 83.9--100\%, 0/20 has 0--16.1\%).}
\label{tab:modalities}
\setlength{\tabcolsep}{4pt}
\begin{tabular}{@{}llcc@{}}
\toprule
Configuration & Active modalities & Candidate & Mission \\
\midrule
Frontal entry       & 4/4 & 20/20 & 20/20 \\
Deep-east entry     & 3/4 (no EO) & 20/20 & 0/20 \\
Rear entry          & 2/4 (no EO, no acoustic) & 20/20 & 0/20 \\
\midrule
All modalities      & 4/4 & 20/20 & 20/20 \\
Radar disabled      & 3/4 & 0/20 & 0/20 \\
RF disabled         & 3/4 & 0/20 & 0/20 \\
EO disabled         & 3/4 & 20/20 & 0/20 \\
Acoustic disabled   & 3/4 & 20/20 & 20/20 \\
\bottomrule
\end{tabular}
\end{table}

\subsection{Dispatch Fan-out (Q4)}

T7 was run in isolation, injecting 10 missions per fleet size $N\in\{1,2,4,8,16\}$. It always produced two tasks per mission (one intercept, one observation), independently of $N$: selecting the nearest UAV scans the fleet, but tasks go only to the selected vehicles. Outgoing traffic per mission is thus constant by design, consistent with the analysis of TB-CSPN~\cite{borghoff2025tbcspn}; a scalability evaluation of the full chain remains to be done.

\subsection{Safety Layer in Isolation (Q5)}

The VSB was exercised in five runs, each feeding the board a sequence of six sensor scenarios under a constant raw command (Table~\ref{tab:vsb}). No coordination topic was present in any run, confirming the logical independence of the board. Under unsafe conditions the raw command was forced to zero in 100\% of samples in four runs; in the run with the slowest reactions, one sample out of 400 escaped the block during a veto episode, which is consistent with the timing jitter of a board that shares the middleware with every other node. All four guards fired in every run, covering both the core and the extended envelope. Initializing the board in the fail-safe state removed a start-up transient observed in an earlier version. Measured reaction times were about 100--190\,ms for proximity and between a few milliseconds and about 90\,ms for the attitude, battery and altitude guards. These are latencies of the simulation loop and DDS transport, not of analog comparators, which react in microseconds.

\begin{table}[t]
\centering
\caption{Virtual Safety Board in isolation (5 runs).}
\label{tab:vsb}
\setlength{\tabcolsep}{4pt}
\begin{tabular}{@{}lc@{}}
\toprule
Property & Outcome \\
\midrule
No coordination topic present          & 5/5 \\
Command blocked under unsafe condition & 100\% in 4/5 runs; 399/400 in 1 \\
Core envelope (proximity, attitude, battery) & 3/3 guards fired, every run \\
Extended envelope (altitude)           & 1/1 guard fired, every run \\
\bottomrule
\end{tabular}
\end{table}

\subsection{Fault Injection (Q6)}
\label{sec:fault}

Four attacks were injected into the integrated twin, each in at least 10 runs driven by scripts with recorded seeds and code version (Table~\ref{tab:fault}).

\emph{Forged mission.} A \tok{MissionToken} is published directly on the dispatch topic, bypassing T6. As expected, the digital layer accepts it: ROS\,2 topics are not authenticated, so T7 dispatches the nearest UAV toward a building. Here the worker is honest and only the bus is compromised. The VSB stopped the UAV in every run, with closest approaches of $6.23\pm0.14$\,m when the honest worker applied the veto and $6.30\pm0.21$\,m when the independent gate did; after braking, the UAV returned to the position it held when the veto fired.

\emph{Compromised worker.} A tampered worker drives the UAV toward a frontal building and ignores the veto. With the veto left to the worker, PX4 received $18.1\pm0.9$ unsafe commands per run and the UAV struck the building at 4.94\,m/s in every run. With the independent gate, no unsafe command reached PX4 and the UAV stopped $5.88\pm0.27$\,m from the building in every run, the veto taking effect within about 95\,ms of the lidar reporting the building closer than 10\,m. The only difference between the two configurations is the gate.

\emph{Spoofed sensing.} With the four sensing consultants switched off, forged radar, RF and EO observations of a non-existent target are injected. The chain fuses them into a confirmed threat in every run, since the digital layer cannot tell genuine from forged evidence, but a mission forms only when the supervisor authorizes it.

\emph{Clock denial.} The VSB's clock is held at zero, the stall that occurs when the simulated clock does not advance. Unable to judge the freshness of its sensor data, the board fails safe: it vetoes every command although all sensors are healthy, and the vehicle stays grounded for the whole run. With a normal clock the vehicle operates in every run, with no command denied. The board fails in the safe direction, but anyone who can manipulate its clock can ground a healthy vehicle at will: this is an attack on availability, not on safety. A bench test of the board in isolation, with a 5\,s staleness limit, shows the opposite failure. With a running clock, sensors that stop publishing are flagged as stale $5.02\pm0.03$\,s later in every run; with the clock frozen at a plausible non-zero value, they are never flagged, and the board keeps trusting their last readings. Value checks do not read the clock and remain effective: with the clock frozen, a proximity reading of 5\,m still triggers a veto in every run, $0.09\pm0.24$\,s after it arrives.

\begin{table}[t]
\centering
\caption{Fault injection (10 runs per configuration; integrated twin except for the bench rows, board in isolation; 0/10 has 95\% CI 0--27.8\%, 10/10 has 72.2--100\%).}
\label{tab:fault}
\setlength{\tabcolsep}{3pt}
\footnotesize
\begin{tabular}{@{}>{\raggedright\arraybackslash}p{15mm}>{\raggedright\arraybackslash}p{19mm}>{\raggedright\arraybackslash}p{47mm}@{}}
\toprule
Attack & Configuration & Outcome \\
\midrule
\multirow{2}{15mm}{Forged mission} & veto by worker & 0/10 collisions; closest $6.23\pm0.14$\,m \\
 & independent gate & 0/10 collisions; closest $6.30\pm0.21$\,m \\
\midrule
\multirow{2}{15mm}{Compromised worker} & veto by worker & 10/10 collisions at 4.94\,m/s; $18.1\pm0.9$ unsafe commands reach PX4 \\
 & independent gate & 0/10 collisions; 0 unsafe commands; stop $5.88\pm0.27$\,m \\
\midrule
\multirow{2}{15mm}{Spoofed sensing} & authorize & threat 10/10; mission 10/10 \\
 & deny & threat 10/10; mission 0/10 \\
\midrule
\multirow{2}{15mm}{Clock denial} & clock held at 0 & grounded 10/10; 100\% of commands denied \\
 & normal clock & operating 10/10; 0\% denied \\
\midrule
\multirow{3}{15mm}{Clock frozen (bench)} & running, sensors stop & stale veto 10/10 after $5.02\pm0.03$\,s \\
 & frozen, sensors stop & no veto 10/10 \\
 & frozen, proximity 5\,m & veto 10/10 after $0.09\pm0.24$\,s \\
\bottomrule
\end{tabular}
\end{table}

\subsection{Authorization Delay and Fusion Window (Q7)}
\label{sec:sweeps}

Two sweeps on the nominal scenario, 20 runs per value, examine the two timing parameters that most directly shape the chain (Table~\ref{tab:sweeps}). A control series on the same code, with the seeds of Q1, gave $11.07\pm0.20$\,s from first observation to candidate and $12.17\pm0.24$\,s to mission (against $11.07\pm0.25$ and $12.16\pm0.25$\,s in Q1), so the changes made for fault injection left the timing of the chain unchanged.

\emph{Authorization delay.} With the supervisor's decision delayed by 0, 5 or 10\,s after the first confirmed threat, a mission is released in every run as soon as the authorization arrives; with 20\,s, never. The boundary follows from token lifetimes and geometry, not from the architecture as such. In the frontal entry, the threat is reconfirmed only while the intruder crosses the EO field, for about 8.5\,s. Since each \tok{ConfirmedThreat} is valid for 10\,s, the threat stays valid for $18.5\pm0.4$\,s after the first confirmation (17.4--19.3\,s over the 80 runs). An authorization arriving later finds no valid threat, and no new confirmation arrives during its own 15\,s validity. This is the intended behavior of governed release: an authorization cannot revive evidence that has expired. It also means that the threat TTL bounds the time available for human deliberation, so that this TTL trades responsiveness against the risk of acting on stale evidence and should be set from the expected decision time, not only from sensor rates.

\emph{Fusion window.} Varying the window of T3 from 0.5 to 4\,s changes neither completion nor the time to candidate (11.03--11.07\,s): radar reports at 5\,Hz and RF at 4\,Hz, so the nearest concurring pair is always less than 0.1\,s apart. The recordings show a secondary effect. A narrower window forms fewer candidates (3.65 per second at 0.5\,s, against 4.5--4.6 at 3--4\,s). Since T5 consumes one candidate per confirmed threat, EO observations queue up and are still fused after the camera has lost the intruder. In the runs in which the EO carrier stays on station, the last confirmed threat follows the last EO observation by $1.77\pm0.60$\,s at 0.5\,s, against about 0.15\,s at 4\,s, and the threat stays valid longer. All of this is within T5's 3\,s tolerance, but it exposes the limitation on evidence age noted in Sec.~\ref{sec:twin-net}.

\begin{table}[t]
\centering
\caption{Sensitivity to authorization delay and T3 window (nominal scenario, 20 runs per value; times in simulation seconds).}
\label{tab:sweeps}
\setlength{\tabcolsep}{4pt}
\footnotesize
\begin{tabular}{@{}llccc@{}}
\toprule
Parameter & Value & Mission & Obs.\ $\to$ mission & Threat valid for \\
\midrule
\multirow{4}{*}{Auth.\ delay} & 0\,s & 20/20 & $12.17\pm0.24$ & $17.98\pm0.48$ \\
 & 5\,s  & 20/20 & $17.16\pm0.26$ & $18.86\pm0.28$ \\
 & 10\,s & 20/20 & $22.17\pm0.27$ & $18.51\pm0.15$ \\
 & 20\,s & 0/20  & -- & $18.59\pm0.15$ \\
\midrule
\multirow{4}{*}{T3 window} & 0.5\,s & 20/20 & $12.11\pm0.25$ & $19.67\pm0.67$ \\
 & 1\,s & 20/20 & $12.19\pm0.25$ & $19.16\pm0.73$ \\
 & 2\,s & 20/20 & $12.12\pm0.25$ & $18.34\pm0.64$ \\
 & 4\,s & 20/20 & $12.13\pm0.22$ & $18.01\pm0.48$ \\
\bottomrule
\end{tabular}\\[2pt]
\footnotesize 20/20: 95\% CI 83.9--100\%; 0/20: 0--16.1\%. Nominal T3 window: 3\,s (first row).
\end{table}

\section{Discussion}
\label{sec:discussion}

\subsection{What Execution Revealed}

\emph{Time is part of the attack surface.} In an early configuration the physical-layer nodes ran on wall-clock time while sensors stamped readings in simulation time. Every reading then looked stale, and the VSB vetoed every command, freezing the fleet. The clock-denial experiment reproduces this failure deliberately: a staleness guard turns any fault of the time base into a loss of availability, so an adversary who can stall the clock of the safety layer can stop the whole swarm. The bench test shows the opposite failure: if the clock is frozen at a plausible non-zero value instead, a sensor that stops updating never looks stale, and the board keeps trusting its last reading, so the guard no longer stops the vehicle when it should. Both failures have one cause: the board measures the age of its data against a clock that an attacker can manipulate. An independent, tamper-resistant time reference, such as a free-running timer on the safety board itself, removes both; the analog channels of the architecture need no time base at all.

\emph{A veto is only as strong as its enforcement point.} The compromised-worker experiment shows that a veto enforced by the component it is meant to restrain is no veto at all: the same board, with the same guards, prevented no collision when the worker applied its verdict and every collision when an independent gate did. In hardware, the enforcement point must lie outside every component that can be compromised, between the flight controller output and the actuators, which is where the analog board of Guarded Swarms places it. The forged-mission and spoofed-sensing experiments also delimit the digital guarantees. TB-CSPN guards protect against faulty or incoherent evidence, not against an adversary who can write to the bus or forge coherent evidence. Against forged evidence, human authorization is the only digital barrier; against a forged mission, only the safety layer remains. Topic authentication such as SROS2~\cite{mayoral2022sros2} would close the first gap on the software side, but it would not remove the need for an enforcement point outside software.

\emph{Governed release requires idempotent consumers.} A first version of T7 deduplicated by mission identifier. Because T6 re-emits a fresh mission at each reconfirmation, every mission looked new, and the nearest UAV was re-selected each time. As the intruder moved, the nearest UAV alternated, both UAVs were tasked to the same target, converged and collided. The VSB was active, but the UAVs' proximity sensor is a single forward-looking lidar ray, and the two vehicles, starting at 12\,m and 20\,m, converged on the same point at the 15\,m intercept altitude, approaching each other from above, below and the side, outside the ray. After the impact the attitude guard fired, but freezing a position setpoint cannot right a capsized multirotor. Deduplicating by \emph{target} fixed the defect: a target is assigned once and the intercepting UAV keeps tracking it. The general lesson is that when authorizations are modeled as non-consumed, time-bounded tokens, every downstream transition must be idempotent with respect to target identity. The specification left this obligation implicit; the twin made it explicit. The episode also marks two requirements for the hardware safety layer: proximity sensing must cover all directions, and the veto must act at the actuators rather than on setpoints.

\emph{Mission validity is not track persistence.} Confirmations also update the threat position estimate. When the intruder leaves the fixed EO cone, confirmations stop, the mission remains valid until its TTL expires, and the interceptor holds the last known position, then releases the mission and returns. The architectural papers treat mission validity and tracking abstractly and together; the twin shows that they are separate properties with separate lifetimes. The same coverage limit explains the directional detection of Table~\ref{tab:modalities}. A steerable camera, or a position estimate that survives brief losses of confirmation, would decouple them.

\subsection{What the Twin Validates and What It Does Not}

Table~\ref{tab:fidelity} summarizes the fidelity boundary. Vehicle dynamics are credible because PX4 SITL runs the real autopilot code. The coordination logic, including guards, windows, governed release and negative guarantees, is software, and the twin exercises it as such under the tested conditions; these are properties of the architecture, not of the simulator. For the safety layer, the twin validates the \emph{logic} (which guards fire, when, with what envelope coverage) and its \emph{logical} independence from coordination. It does not validate physical independence: the VSB runs on the same machine and middleware as the coordination nodes, acts on position setpoints rather than cutting PWM at the actuator, and therefore cannot restore attitude once a vehicle is already compromised. Resistance to a compromise of the whole computing platform, which is the central claim of Guarded Swarms, can only be demonstrated on separate hardware. Sensing is synthesized from ground truth through coverage, confidence and noise profiles, so the twin validates the structure of fusion, not detector accuracy.

\begin{table}[t]
\centering
\caption{Fidelity boundary of the twin.}
\label{tab:fidelity}
\setlength{\tabcolsep}{3pt}
\begin{tabular}{@{}>{\raggedright\arraybackslash}p{22mm}>{\raggedright\arraybackslash}p{29mm}>{\raggedright\arraybackslash}p{28mm}@{}}
\toprule
Aspect & Validated in twin & Requires hardware \\
\midrule
Vehicle dynamics & credible (PX4 SITL) & flight tests \\
Coordination logic & guards, windows, release, negative guarantees & real networks, contested links \\
Safety-layer logic & guards, coverage, logical independence & physical independence, \si{\micro\second} reaction, PWM cut \\
Sensing & fusion structure & detector accuracy \\
Human oversight & no action before authorization & real deliberation, HMI \\
\bottomrule
\end{tabular}
\end{table}

\subsection{Relevance to Responsible Military Autonomy}

The executable chain gives concrete form to several NATO principles of responsible use~\cite{nato2021ai}. \emph{Governability} is enforced twice: by T6, without which no mission exists, and by the VSB, which no mission can override. \emph{Traceability} is provided by the provenance chain from command to observation and to the authorizing supervisor. \emph{Reliability} is supported by negative guarantees that hold structurally at three distinct stages. The twin also makes the limits of these properties measurable: authorization guarantees ordering but not timeliness, and confirmation is only as good as the narrowest sensor.

\subsection{Threats to Validity}

The coordination experiments (Q1--Q3) use 20 runs per configuration, which bounds an unobserved failure rate only to about 16\% at 95\% confidence; the fault-injection experiments use 10 runs per configuration (bound of about 28\%), and the dispatch and safety-layer experiments still rest on smaller runs (10 and 5). All runs come from a single scenario on a single machine. Authorization is issued by a scripted supervisor node, so the human component of latency is not measured. Sensing is synthetic. The dispatch experiment isolates one transition. An offline audit of an earlier set of recordings, reconstructing every instant in simulation time from the recorded clock, proved essential: it showed that the test harness, not the architecture, caused an incomplete run and extreme latencies that we had initially attributed to the timing of EO confirmations. Harness effects can masquerade as architectural behavior, and twins used for TEV\&V need the same audit discipline as the systems they test; the campaign reported here therefore checks every run automatically before it enters the analysis. The negative-guarantee and safety-layer results are conformance tests of an implementation against its specification; they show that the implementation realizes the guarantees under the tested conditions, not that the guarantees hold in general, which is the role of the formal TB-CSPN analysis~\cite{borghoff2025tbcspn}.

\section{Conclusion and Future Work}

We have turned Guarded Swarms from an architectural description into a running, measurable system. The TB-CSPN chain forms missions from multi-modal evidence under human authorization, refuses to form them from conflicting, stale or unauthorized evidence at three distinct stages, and exposes the dependence of threat confirmation on the narrowest modality. The Virtual Safety Board enforces its modeled envelope independently of coordination logic, and under fault injection it prevented every collision caused by a forged mission or a compromised worker once its veto was enforced by an independent gate, while a veto left to the compromised worker prevented none. Executing the architecture also surfaced obligations that the specification left implicit (tamper-resistant time bases, enforcement points outside every component that can be compromised, idempotent consumers of governed release, evidence age carried through every fused token, and the separation of mission validity from track persistence) and drew a sharp line between what simulation can establish and what only hardware can.

Three directions follow. 
First, steerable or redundant EO coverage, to make confirmation independent of the entry vector. 
Second, a supervision console, to replace scripted authorization with real human decisions, enabling measures of authorization latency and operator workload. 
Third, and most important, a physical safety board with the channels not modeled here (thrust clamping, power and thermal protection, RF guard tone, emergency stop), along the hardware path of~\cite{borghoff2026fpga}, to test the guarantees that the twin can only approximate.
\balance

\section*{Acknowledgment}
The work of R.~Pareschi was partially supported by the ERMETE project (Extended Reality for MaintenancE and Training Ecosystem), funded by the Italian Ministry of Enterprises and Made in Italy (MIMIT), grant B39J25000400005.



\end{document}